\documentclass[prd,eqsecnum,twocolumn,amsfonts,amssymb]{revtex4}

\usepackage{graphicx}

\usepackage{bm}

\newcommand{\beq}{\begin{equation}}
\newcommand{\eeq}{\end{equation}}
\newcommand{\beqs}{\begin{eqnarray}}\newcommand{\eeqs}{\end{eqnarray}}

\newcommand{\gsim}{\mathrel{\raisebox{-
.6ex}{$\stackrel{\textstyle>}{\sim}$}}}

\begin{document}

\title{Nucleon Decay and $n-\bar n$ Oscillations in a Left-Right Symmetric
  Model with Large Extra Dimensions} 

\author{Sudhakantha Girmohanta and Robert Shrock}

\affiliation{ \ C. N. Yang Institute for Theoretical Physics and 
Department of Physics and Astronomy, \\
Stony Brook University, Stony Brook, NY 11794, USA }

\begin{abstract}

We study baryon-number-violating processes, including proton and bound
neutron decays and $n-\bar n$ oscillations, in a left-right-symmetric
(LRS) model in which quarks and leptons have localized wavefunctions
in extra dimensions. In this model we show that, while one can easily
suppress baryon-number-violating nucleon decays well below
experimental bounds, this does not suppress $n-\bar n$ transitions,
which may occur at levels comparable to current limits.  This is
qualitatively similar to what was found in an extra-dimensional model
with a Standard-Model low-energy effective field theory (SMEFT).  We
show that experimental data imply a lower limit on the mass scale
$M_{n \bar n}$ characterizing the physics responsible for $n-\bar n$
oscillations in the LRS model that is significantly higher than in the
extra-dimensional model using a SMEFT and explain the reason for this.
Our results provide further motivation for new experiments to search
for $n-\bar n$ oscillations.

\end{abstract}

\maketitle


\section{Introduction}
\label{intro_section}

The Standard Model (SM) conserves baryon number, $B$
\cite{su2instantons,hooft}, but baryon-number violation (BNV) is
expected to occur in nature, since this is one of the requisite
conditions for producing the observed baryon number asymmetry in the
universe \cite{sakharov}. Indeed, many ultraviolet extensions of the
Standard Model, such as grand unified theories (GUTs), do feature
baryon-number violation (as well as the violation of total lepton
number, $L$). In addition to the $\Delta B=-1$ decays of protons and
bound neutrons, another type of baryon number violation is
neutron-antineutron oscillations, with $|\Delta B|=2$. These $n-\bar
n$ oscillations could explain baryogenesis \cite{kuzmin}. Some early
studies of $n-\bar n$ oscillations include
\cite{glashow}-\cite{nnb84}. The same physics beyond the Standard
Model (BSM) that gives rise to $n-\bar n$ oscillations also leads to
matter instability via $\Delta B=-2$ decays of $nn$ and $np$ dinucleon
states in nuclei.  Several generations of experiments have searched
for baryon-number-violating decays of protons and bound neutrons
(henceforth denoted simply as nucleon decays) and have set limits on
such decays \cite{pdg}.  There have also been searches for $n-\bar n$
oscillations using neutron beams from reactors \cite{ill} and for
matter instability and various dinucleon decay modes using large
underground detectors \cite{pdg}. The best current limit on matter
instability is from the Super-Kamiokande (SK) experiment
\cite{sk_nnbar}.

The operators in the low-energy effective Hamiltonian (in four
spacetime dimensions) for proton decay are four-fermion operators with
Maxwellian (i.e., free-field) mass dimension 6 and hence coefficients
of mass dimension $-2$, whereas the operators in ${\cal H}^{(n \bar
  n)}_{eff}$ are six-quark operators, with coefficients of dimension
$-5$.  Hence, if there were only a single mass scale characterizing
BNV physics, then nucleon decays would generically be much more
important as a manifestation of baryon number violation than
$n-\bar n$ oscillations and the corresponding dinucleon decays.  However, the
opposite order of importance of BNV processes may actually describe
nature.  In Ref. \cite{mm80}, Mohapatra and Marshak presented a model
using a left-right symmetric gauge group (in four spacetime
dimensions) in which $n-\bar n$ oscillations occur, while proton decay
does not. In Ref. \cite{nnb02}, Nussinov and Shrock presented an
extra-dimensional model in which proton decay is suppressed well
beyond observable levels while $n-\bar n$ oscillations occur at levels
comparable to experimental limits. In the model used in \cite{nnb02},
quarks and leptons having strongly localized wavefunction profiles in
the extra dimensions \cite{as,ms}.  In the models of both
Refs. \cite{mm80} and \cite{nnb02}, it is the $n-\bar n$ oscillations
and the corresponding $nn$ and $np$ dinucleon decays to multi-meson
final states that are the main manifestations of baryon number
violation, rather than individual BNV nucleon decays.  Further
examples of models in four spacetime dimensions with baryon number
violation but no proton decay were later given in
\cite{wise}. Recently, in \cite{bvd} we studied a number of related
BNV nucleon and dinucleon decays to various final states in the
extra-dimensional model used in \cite{nnb02}.

In this paper we investigate nucleon decays and $n-\bar n$ oscillations
in an extra-dimensional model with the left-right symmetric (LRS) gauge group
\beq
G_{LRS} = {\rm SU}(3)_c \otimes {\rm SU}(2)_L \otimes {\rm SU}(2)_R \otimes 
{\rm U}(1)_{B-L} \ . 
\label{glrs}
\eeq
Our present work complements the study in Ref.  \cite{mm80}, which was
set in four spacetime dimensions, and also the previous studies
\cite{nnb02} and \cite{bvd}, which used a low-energy effective field
theory with the SM gauge group, $G_{SM} = {\rm SU}(3)_c \otimes {\rm
  SU}(2)_L \otimes {\rm U}(1)_Y$ rather than $G_{LRS}$. Anticipating
our results in advance, we show that in the extra-dimensional LRS
model, it is easy to suppress nucleon decays well below observable
levels, but this does not suppress $n-\bar n$ oscillations, which can
occur at levels comparable with current experimental limits.  This is
qualitatively similar to the conclusions reached in \cite{nnb02}.
Here we find an interesting feature of the extra-dimensional LRS model
that makes $n-\bar n$ oscillations even less suppressed than in the
model of \cite{nnb02} with its SMEFT. The reason for this is that the
integration of six-quark operators over the extra dimensions always
led to exponential suppression factors in the model of \cite{nnb02},
whereas, in contrast, we find that in the LRS model, there are some
operators for which this integration does not lead to exponential
suppression factors.

Our work here also complements our recent studies in \cite{ndl}, where
we derived improved upper bounds on the rates for several
nucleon-to-trilepton decay modes and in \cite{dnd}, where we presented
improved upper bounds on the rates for several dinucleon-to-dilepton
decay channels (see also \cite{ndnd}).  Refs.  \cite{ndl,dnd} were
model-independent phenomenological analyses, whereas our present paper
is a study within the context of a specific type of extra-dimensional
model. Recent reviews of $n-\bar n$ oscillations include
\cite{mohapatra_rev,nnbar_physrep}.

This paper is organized as follows.  In Sec. \ref{lrs_section} we briefly
review the properties of the left-right symmetric model that will be needed for
our analysis. In Sec. \ref{extra_dimension_section} we discuss the
extra-dimensional model and low-energy effective field theory approach that
serve as the theoretical framework for our calculations. In
Sec. \ref{pdecay_section} we extract constraints on the fermion wavefunctions
in the model from limits on BNV nucleon decay modes.  Section
\ref{nnbar_section} contains our analysis of $n - \bar n$ oscillations.  Our
conclusions are presented in Section \ref{conclusion_section}. 


\section{Left-Right Symmetric Model}
\label{lrs_section} 

In this section we recall some basic properties of the left-right 
symmetric model \cite{mm80},\cite{lrs75a}-\cite{lrs81} that will be relevant
here, and define our notation for the fermion
and Higgs fields in the theory. The Lagrangian is invariant under the gauge
group $G_{LRS}$ in Eq. (\ref{glrs}), with corresponding SU(2)$_L$, SU(2)$_R$,
and U(1)$_{B-L}$ gauge fields ${\vec A}_{L,\mu}$, ${\vec A}_{R,\mu}$ and
$U_\mu$, and respective gauge couplings $g_L$, $g_R$, and $g_U$. The quarks and
leptons of each generation transform as
\beq
Q_L: \ (3,2,1)_{1/3,L} \ , \quad Q_R: \ (3,1,2)_{1/3,R}
\label{quarks}
\eeq
and
\beq
L_{\ell,L}: \ (1,2,1)_{-1,L} \ , \quad L_{\ell,R}: \ (1,1,2)_{-1,R} \ , 
\label{leptons}
\eeq
where the numbers in the parentheses are the dimensionalities of the
representations under the three non-Abelian factor groups in $G_{LRS}$ and the
numbers in the subscripts are the values of $B-L$. (No confusion should result
from the use of the symbol $L$ for both ``lepton'' and ``left''; the context
will make clear which meaning is intended.) For our purposes, we shall only
need the first-generation quark fields, which are, explicitly,
\beq
Q^\alpha_L = {u^\alpha \choose d^\alpha}_L \ , \quad 
Q^\alpha_R = {u^\alpha \choose d^\alpha}_R \ , 
\label{qud}
\eeq
where Greek indices $\alpha, \ \beta$, etc. are SU(3)$_c$ color indices. The
explicit lepton field are 
\beq
L_{\ell,L} = {\nu_\ell \choose \ell}_L \ , \quad 
L_{\ell,R} = {\nu_\ell \choose \ell}_R \ , 
\label{explicitleptons}
\eeq
where $\ell=e, \ \mu, \tau$. We denote SU(2)$_L$ and SU(2)$_R$ gauge indices
as Roman indices $i,j..$ and primed Roman indices $i',j'...$, respectively, so,
e.g., $Q^{i \alpha}_L = u^\alpha_L$ for $i=1$ and 
$Q^{i' \alpha}_R = d^\alpha_R$ for $i'=2$.  The electric charge is given by the
elegant expression $Q_{em} = T_{3L}+T_{3R}+(B-L)/2$, where $\vec T_L$ and 
$\vec T_R$ denote the SU(2)$_L$ and SU(2)$_R$ weak isospin generators. 

The Higgs sector contains a Higgs 
field $\Phi$ transforming as $(1,2,2)_0$, which 
can be written as $\Phi^{i j'}$, or equivalently, in matrix form, as
\beq
\Phi = \left( \begin{array}{cc}
    \phi_1^0 & \phi_1^+ \\
    \phi_2^- & \phi_2^0 \end{array} \right ) \ . 
\label{phi}
\eeq
The Higgs sector also contains two Higgs fields, commonly denoted $\Delta_L$ 
and $\Delta_R$, which transform as $(1,3,1)_2$ and $(1,1,3)_2$, respectively. 
Since the adjoint representation of SU(2) is equivalent to the symmetric rank-2
tensor representation, these may be written as $(\Delta_L)^{ij} = 
(\Delta_L)^{ji}$ and $(\Delta_R)^{i'j'} = (\Delta_R)^{j'i'}$ or, 
alternatively, as (traceless) matrices:
\beq
\Delta_\chi = \left( \begin{array}{cc}
    \Delta_\chi^+/\sqrt{2} & \Delta_\chi^{++} \\
    \Delta_\chi^0 & -\Delta_\chi^+/\sqrt{2} \end{array} \right ) \ , 
\quad \chi=L, \ R .
\label{Delta}
\eeq
The minimization of the Higgs potential to produce vacuum expectation values
(VEVs) has been analyzed in a number of studies
\cite{lrs81},\cite{dgko}-\cite{dmrx}.  With appropriate choices of parameters
in the Higgs potential, this minimization yields the following vacuum
expectation values (VEVs) of the Higgs fields:
\beq
\langle \Phi \rangle_0 = \frac{1}{\sqrt{2}} \left( \begin{array}{cc}
    \kappa_1 &  0 \\
     0       & \kappa_2 e^{i\theta_\Phi} \end{array} \right ) \ , 
\label{phivev}
\eeq
\beq
\langle \Delta_L \rangle_0 = \frac{1}{\sqrt{2}} \left( \begin{array}{cc}
     0       &  0 \\
     v_L e^{i\theta_{\Delta}} & 0 \end{array} \right ) \ 
\label{deltaLvev}
\eeq
and
\beq
\langle \Delta_R \rangle_0 = \frac{1}{\sqrt{2}} \left( \begin{array}{cc}
            0   &  0 \\
            v_R &  0  \end{array} \right ) \ . 
\label{deltaRvev}
\eeq
(Here, the choices of which VEVs are real are made with the requisite
rephasings.) The spontaneous symmetry breaking of the $G_{LRS}$ gauge symmetry
occurs in several stages.  At the highest-mass stage, $\Delta_R$ picks up a
VEV, thereby breaking the ${\rm SU}(2)_R \otimes {\rm U}(1)_{B-L}$ subgroup of
$G_{LRS}$ to U(1)$_Y$, where $Y$ denotes the weak hypercharge, i.e.,
\beq
    {\rm SU}(2)_R \otimes {\rm U}(1)_{B-L} \to {\rm U}(1)_Y \ .
\label{symbreak1}
\eeq
This gives the $W_R$ a large mass, which, to leading order, is
$m_{W_R} = g_Rv_R/\sqrt{2}$. The second stage of symmetry breaking, 
\beq
{\rm SU}(2)_L \otimes {\rm U}(1)_Y \to {\rm U}(1)_{em} \ , 
\label{symbreak2}
\eeq
occurs at a lower scale and results from the the VEVs of the $\Phi$ field.
This gives a mass $m_{W_L} = g_L v_{EW}/2$, where $v_{EW} = \sqrt{\kappa_1^2 +
  \kappa_2^2} = 246$ GeV is the electroweak symmetry breaking (EWSB) scale.
The neutral gauge fields $A_{3L}$, $A_{3R}$, and $U$ mix to form the photon,
the $Z$, and a much more massive $Z'$.  Since the VEV $v_L$ of the SU(2)$_L$
Higgs triplet $\Delta_L$ would modify the successful tree-level relation $\rho
= 1$, where $\rho = m_W^2/(m_Z^2\cos^2\theta_W) = 1$, one takes $v_L \ll
\kappa_{1,2}$. It is also possible to consider dynamical breaking of the LRS
gauge symmetry (e.g., \cite{dynamical_lrs,sml}), but the conventional scenario
with Higgs fields will be assumed here.

This LRS model has several interesting features as a UV extension of
the Standard Model. The relation for $Q_{em}$ entails charge
quantization. Furthermore, one may impose left-right symmetry at some
high ultraviolet (UV) scale, so the running gauge couplings for
SU(2)$_L$ and SU(2)$_R$ are equal, i.e., $g_L = g_R$ at this scale,
thereby reducing the number of parameters in the model.  The
left-right symmetry in the Lagrangian is of conceptual interest since
it means that parity violation is due to spontaneous symmetry
breaking, rather than being intrinsic, as in the Standard Model. The
non-observation of any right-handed charged currents in weak decays
and the lower limits (of order several TeV) from the Large Hadron
Collider on a $W_R^\pm$ and $Z'$ can be accommodated by making $v_R$
sufficiently large. Since the $\Delta_R$ has $B-L$ charge of 2, its
VEV, $v_R$, breaks $B-L$ by two units. The gauge group $G_{LRS}$ has a
natural UV extension to a theory with gauge group $G_{422} = {\rm
  SU}(4)_{PS} \otimes {\rm SU}(2)_L \otimes {\rm SU}(2)_R$, where the
Pati-Salam (PS) gauge group SU(4)$_{PS}$ \cite{ps} contains ${\rm
  SU}(3)_c \otimes U(1)_{B-L}$ as a maximal subgroup. In turn,
$G_{422}$ is a maximal subgroup of the SO(10) GUT group, since ${\rm
  SO}(10) \supseteq {\rm SO}(6) \otimes {\rm SO}(4) \approx {\rm
  SU}(4) \otimes {\rm SU}(2) \otimes {\rm SU}(2)$. There are also
supersymmetric extensions of the LRS model (e.g., \cite{susylrs}).
However, since the LHC has not yet observed evidence of supersymmetric
partners, and since we use a low-energy effective field theory
framework for our analysis, the non-supersymmetric version of the LRS
model will be sufficient for our study.


\section{Extra-Dimensional Framework}
\label{extra_dimension_section} 

In this section we describe the extra-dimensional model that we use.
Some aspects of this discussion are similar to those of
Refs. \cite{nnb02,bvd}, but to make our presentation self-contained,
we reiterate these here. The general motivation for considering extra
(spatial) dimensions dates back to the work of Kaluza \cite{kaluza}
and Klein \cite{klein}, and was considerably strengthened with the
development of string theory as a theory of quantum gravity. The
particular type of extra-dimensional model that was used for the study
of $n-\bar n$ oscillations in \cite{nnb02,bvd} has the appeal that it
can naturally explain the large hierarchy in quark and lepton masses
by requisite properties of fermion wavefunctions in the extra
dimensions, without the need for a large range of dimensionless Yukawa
couplings in the fundamental theory \cite{as,ms}.

A remark is in order concerning a difference in our use of the
extra-dimensional model here and the use in Refs. \cite{nnb02} and
\cite{bvd}.  Because the scale of baryon-number violation responsible
for $n-\bar n$ oscillations is larger than the electroweak scale,
Refs. \cite{nnb02} and \cite{bvd} used a low-energy effective field
theory analysis with six-quark operators that are invariant under the
Standard Model gauge group, $G_{SM}$, i.e., an extra-dimension
SMEFT. As noted above, in the Standard Model, $B$ is a global
symmetry, and the baryon-number-violating physics that gives rise to
$n-\bar n$ oscillations is encoded in the six-quark operators and
their coefficients.  In contrast, in the LRS model, $B$ and $L$ are
both gauged, as the combination $B-L$ in the U(1)$_{B-L}$ factor group
of $G_{LRS}$.  This gauge symmetry is spontaneously broken by the VEV
of the $\Delta_R$ field at the high scale $v_R$. As mentioned above,
since $\Delta_R$ has charge 2 under U(1)$_{B-L}$, this VEV $v_R$
breaks U(1)$_{B-L}$ by two units.  For a process that has $\Delta
L=0$, this means that it breaks $B$ as $|\Delta B|=2$. It follows that
the mass scale, $M_{n \bar n}$, characterizing the physics responsible
for $n-\bar n$ oscillations is $v_R$:
\beq
M_{n \bar n}=v_R \ . 
\label{m_nnb_vr}
\eeq
We shall analyze $n-\bar n$ oscillations in this
theory by writing down the relevant $G_{LRS}$-invariant operators, which
are six-quark operators multiplied by $(\Delta_R)^\dagger$, and then
focusing on the resultant six-quark operators resulting from the VEV of
$(\Delta_R)^\dagger$. 

Proceeding with the description of the extra-dimensional model, the usual
spacetime coordinates are denoted as $x_\nu$, with $\nu=0,1,2,3$, and the $n$
extra coordinates as $y_\lambda$ with $1 \le \lambda \le n$; 
for definiteness, the latter are assumed to
be compact. The fermion and boson fields are taken to have a factorized form.
For fermions, this form is
\beq
\Psi(x,y)=\psi(x)\chi(y) \ , 
\label{psiform}
\eeq
where here $\Psi(x,y)$ is a generic symbol standing for $Q_L(x,y)$,
$Q_R(x,y)$, $L_{\ell,L}(x,y)$ or $L_{\ell,R}(x,y)$. In the extra dimensions
these fields are restricted to the interval $0 \le y_\lambda \le L$
for all $\lambda$. We define an energy corresponding to the inverse
of the compactification scale as $\Lambda_L \equiv 1/L$.  

Starting from an effective Lagrangian in the $d=(4+n)$-dimensional spacetime,
one obtains the resultant low-energy effective Lagrangian in four dimensions by
integrating over the extra $n$ dimensions.  We use a low-energy effective field
theory (EFT) approach that entails an ultraviolet cutoff, which we denote as
$M_*$.  In accordance with this low-energy EFT approach, as in Ref. \cite{ms},
we focus on the lowest KK modes of the boson (gauge and Higgs) fields and take
these to have flat profiles in the extra dimensions. Recall that the Maxwellian
mass dimension of a boson field in a $d=4+n$ dimensional spacetime is $d_b
=(d-2)/2 = 1+(n/2)$.  Therefore, in order to maintain canonical normalization
of boson fields in four spacetime dimensions, a Higgs field in $4+n$ dimensions
with a flat profile in the extra dimensions, generically denoted $\phi_{4+n}$,
has the form
\beq
\phi_{4+n}(x,y)=(\Lambda_L)^{n/2} \, \phi(x) = L^{-n/2}\phi(x) \ .
\label{phi4pn}
\eeq
It is readily seen that the integration of the quadratic terms in the Higgs
field over the $n$ extra dimensions yields the correct normalization for the
resultant quadratic terms in the Lagrangian in four spacetime dimensions:
\beq
\int_0^L d^n y \, {\rm Tr}[\phi_{4+n}^\dagger \phi_{4+n}] = 
L^n [L^{-n} {\rm Tr}(\phi^\dagger \phi)] = {\rm Tr}(\phi^\dagger \phi) .
\label{phiphi_int}
\eeq
The coefficients of higher-power products of Higgs fields can be
expressed using similar methods.  For example, the coefficient
$\lambda_{1,4+n}$ of the quartic term $[{\rm Tr}(\Phi(x,y)^\dagger
  \Phi(x,y))]^2$ has dimensions $d_{\lambda_{1,4+n}} = 4-d = -n$, and
hence we set $\lambda_{1,4+n} = \Lambda_L^{-n} \lambda_1 = L^n
\lambda_1$ so that the integration over the extra dimensions yields
the standard quartic term in the Lagrangian:
\beqs
&& \lambda_{1,4+n} \int_0^L d^ny \, 
    [{\rm Tr}(\Phi(x,y)^\dagger \Phi(x,y))]^2 \cr\cr
    &=& 
(L^n \lambda_1)(L^n)(L^{-n/2})^4 {\rm Tr}(\Phi(x)^\dagger \Phi(x))] \cr\cr
  &=& \lambda_1 [{\rm Tr}(\Phi(x)^\dagger \Phi(x))] \ , \cr\cr
  &&
\label{phiquartic}
\eeqs
and similarly with other terms in the Higgs potential. Corresponding
statements apply for the covariant derivative terms.  The VEV of the
higher-dimensional Higgs field $(\Delta_R)_{4+n}$ is thus
\beq
\langle (\Delta_R)_{4+n} \rangle_0 = (\Lambda_L)^{n/2} \, v_R = L^{-n/2} \, v_R \ .
\label{vr4pn}
\eeq
Since the Higgs fields are taken to have flat profiles in the extra
dimensions as in \cite{ms} and since we will only need to make use of
their VEVs for our purposes, we may simply replace the various Higgs
fields by these VEVs in the four-spacetime-dimensional Lagrangian and
deal only with the dependence of the fermion fields on the $y$
coordinates.  This simplified procedure will be followed henceforth.

The localization of
the wavefunction of a fermion $f$ in the extra dimensions has the form
\cite{as,ms} 
\beq
\chi_f(y) = A \, e^{-\mu^2 \, \| y-y_f \|^2} \ , 
\label{gaussian}
\eeq
where $A$ is a normalization factor and $y_f \in {\mathbb R}^n$ denotes 
the position vector of this fermion in the extra dimensions, with components 
$y_f = (y_{f,1},...,y_{f,n})$ and with the standard Euclidean norm 
of a vector in ${\mathbb R}^n$, namely 
$\| y_f \| \equiv \Big (\sum_{\lambda=1}^n y_{f,\lambda}^2 \Big )^{1/2}$.
For $n=1$ or $n=2$, this fermion localization can result from appropriate
coupling to a scalar localizer field with a kink or vortex solution,
respectively \cite{rubakov83}-\cite{volkas2007}. Corrections due to Coulombic
gauge interactions between fermions have been studied in \cite{qlw}. The
normalization factor $A$ is determined by the condition that, after integration
over the $n$ higher dimensions, the four-dimensional fermion kinetic term has
its canonical normalization. This yields the result
\beq
A=\bigg ( \frac{2}{\pi} \bigg )^{n/4}\, \mu^{n/2} \ . 
\label{a}
\eeq
We define a distance inverse to the localization measure $\mu$ as
$L_\mu \equiv 1/\mu$. The fermion wavefunctions are assumed to be
strongly localized, with half-width $L_\mu \ll L$ at various points in
the higher-dimensional space.  We define $\xi \equiv L/L_\mu =
\mu/\Lambda_L$. As in the earlier works \cite{nnb02,bvd}, the choice
$\xi \sim 30$ is made for sufficient
separation of the various fermion wavefunctions while still fitting
well within the size $L$ of the compactified extra dimensions. The UV
cutoff $M_*$ is taken to be much larger than any mass scale in the model, to
ensure the self-consistency of the low-energy effective 
field theory analysis. The choice $\Lambda_L \gsim 100$ TeV is
consistent with bounds on extra dimensions from precision electroweak
constraints, and collider searches \cite{pdg} and produces adequate
suppression of flavor-changing neutral-current (FCNC) processes
\cite{dpq2000} (see also \cite{acd,abpy}). With $\xi=30$, this
yields $\mu \sim 3 \times 10^3$ TeV.
(The models
considered here with SM fields propagating in the large extra
dimensions, are to be contrasted with models in which only the
gravitons propagate in these dimensions (e.g.,
\cite{add}-\cite{comp}) and models with noncompact extra dimensions and 
a warped metric \cite{rs1,rs2}.) 

For integrals of products of fermion fields, although the range of 
integration over each of the $n$ coordinates of a vector $y$ is from
0 to $L$, the strong localization of each fermion field in the Gaussian
form (\ref{gaussian}) means that, to a very good approximation, the
restriction of the fermion wavefunctions to the form (\ref{gaussian}),
the range of integration can be extended to the interval
$(-\infty,\infty)$: $\int_0^L d^n y \to \int_{-\infty}^\infty d^n y$.
We define the (dimensionless) vector
\beq
\eta = \mu y \ . 
\label{eta}
\eeq

We next discuss the Yukawa terms and resultant mass terms for quarks in this
extra-dimensional LRS model. These are 
\beq
 {\cal L}_{Yuk} = \sum_{a,b=1}^3 [ \bar Q_{a,L} (y^{(q)}_{ab} \Phi + 
h^{(q)}_{ab} \tilde \Phi) Q_{b,R}]  + h.c. \ ,
\label{yukterm}
\eeq
where $a, \ b$ are generation indices and $\tilde \Phi = \tau_2 \Phi^* \tau_2$,
and here $y^{(q)}_{ab}$ and $h^{(q)}_{ab}$ are Yukawa couplings. 
Inserting the VEV of $\Phi$ from Eq. (\ref{phivev}) and performing the
integration, over the extra dimensions, of the quark bilinears gives the mass
terms
\beqs
&& \frac{1}{\sqrt{2}}\sum_{a,b=1}^3 
[\bar u_{a,L}( y^{(q)}_{ab} \kappa_1 + 
h^{(q)}_{ab} \kappa_2 e^{i \theta_\Phi}) u_{b,R}]\, e^{-S_{yQ,ab}} 
+ \cr\cr
&& 
\frac{1}{\sqrt{2}}\sum_{a,b=1}^3 
[\bar d_{a,L}( y^{(q)}_{ab} \kappa_2 e^{-i\theta_\Phi} 
  + h^{(q)}_{ab} \kappa_1)d_{b,R}] \, e^{-S_{yQ,ab}} + h.c., \cr\cr
&&
\label{quarkmasses}
\eeqs
where
\beq
S_{yQ,ab} = 
\frac{1}{2}\|\eta_{Q_{a,L}} - \eta_{Q_{b,R}} \|^2 \ .
\label{sq}
\eeq
For our study of $n-\bar n$ oscillations in this model, we will only
need to deal with the first-generation quark fields, $Q_{1,L}$ and
$Q_{1,R}$. Consequently, we will omit the generation indices on these
fields, with the understanding that they are first-generation quarks:
$Q_L = {u \choose d}_L$ and .  $Q_R = {u \choose d}_R$. Neglecting
small Cabibbo-Kobayashi-Maskawa mixings, the relevant quark mass terms
are then
\begin{widetext}
\beq
\frac{1}{\sqrt{2}} \, 
\bigg [ [\bar u_L( y^{(q)}_{11} \kappa_1 + h^{(q)}_{11} \kappa_2 e^{i \theta_\Phi}) u_R] + 
 \frac{1}{\sqrt{2}} \, [\bar d_L( y^{(q)}_{11} \kappa_2 e^{-i\theta_\Phi} 
   + h^{(q)}_{11} \kappa_1) d_R] \bigg \} \bigg ]e^{-(1/2)\|\eta_{Q_L}-\eta_{Q_R} \|^2} + h.c.
\label{quarkmasses2}
\eeq
\end{widetext}
Note that although one may impose left-right symmetry in the deep UV, this
symmetry is broken at the scale $v_R$, so 
at this EWSB scale, $\eta_{Q_L}$ is expected to be different from
$\eta_{Q_R}$. In accordance with the original motivation for this type of
extra-dimensional model, namely that the generational hierarchy in the quark
and charged lepton masses is not due primarily to a hierarchy in the
dimensionless Yukawa couplings, but instead to the different positions of the
wavefunction centers in the extra dimensions, one may take 
$y_{11}^{(q)} \sim O(1)$ and
$h_{11}^{(q)} \sim O(1)$. Then
\beq
\|\eta_{Q_L} - \eta_{Q_R}\| = 
\Bigg [ 2 \ln \Bigg ( \frac{| y^{(q)}_{11} \kappa_1 + h^{(q)}_{11}
\kappa_2 e^{i\theta_\Phi}|}{\sqrt{2} \, m_u} \Bigg ) \Bigg ]^{1/2}
\label{qlqrdistance_mu}
\eeq
and
\beq
\|\eta_{Q_L} - \eta_{Q_R}\| = 
\Bigg [ 2 \ln \Bigg ( \frac{| y^{(q)}_{11} \kappa_2 e^{-i\theta_\Phi}
 + h^{(q)}_{11}\kappa_1|}
{\sqrt{2} \, m_d} \Bigg ) \Bigg ]^{1/2}
\label{qlqrdistance_md}
\eeq
For given $\kappa_1$ and $\kappa_2$, the two Yukawa couplings $y^{(q)}_{11}$
and $h^{(q)}_{11}$, and the phase factor $e^{i\theta_\Phi}$ can be chosen to
satisfy these relations. Taking $y_{11}^{(q)} \sim O(1)$ and
$h_{11}^{(q)} \sim O(1)$ as above, and using the values of the running quark masses $m_u$ and
$m_d$ at the EWSB scale from Ref. \cite{koide}, one can then compute a value of
$\|\eta_{Q_L}-\eta_{Q_R}\|$ that satisfies Eqs.  (\ref{qlqrdistance_mu}) and
(\ref{qlqrdistance_md}). For our purposes, we will take the value
\beq
\|\eta_{Q_L}-\eta_{Q_R}\| \simeq 4.7 \ .
\label{distance_QL_Qr}
\eeq

For our analysis of baryon-number-violating processes, let us consider
a generic operator product of fermion fields in the four-dimensional
Lagrangian consisting of $k$ fermion fields multiplied by a
coefficient $c_{r,k}$, which we denote as ${\cal O}_{r,k}$.  We denote
the corresponding operator in the $d=(4+n)$-dimensional space as
$O_{r,k}(x,y)$.  The coefficient of this operator, $\kappa_{r,k}$, can
be written in a form that exhibits its mass dimension explicitly,
namely
\beq
\kappa_{r,k} = \frac{\bar\kappa_{r,k}}
{(M_{BNV})^{k(3+n)/2-4-n}} \ , 
\label{kappagen}
\eeq
where $\bar\kappa_{r,k}$ is dimensionless and $M_{BNV}$ is a relevant
mass scale for the BNV process (nucleon decay or $n-\bar n$ oscillations). 
We denote the integral over the extra dimensions of this fermion operator
product as $I_{r,k}$. Using Eq. (\ref{intform}), we have  
$I_{r,k} = b_k \, e^{-S_{r,k}}$, where 
\beqs
b_k &=& A^k \, \mu^{-n}\Big ( \frac{\pi}{k} \Big )^{n/2} \cr\cr
&=& \Big [ 2^{k/4} \, \pi^{-(k-2)/4} \, k^{-1/2} \, \mu^{(k-2)/2} \Big ]^n \ .
\label{bk}
\eeqs
Then, as in \cite{bvd}, 
\beqs
c_{r,k} &=& \kappa_{r,k} I_{r,k}
 =  \frac{\bar\kappa_{r,k}}{(M_{BNV})^{(3k-8)/2} } \,
 \Big ( \frac{\mu}{M_{BNV}} \Big ) ^{(k-2)n/2} \times \cr\cr
 &\times& 
\bigg ( \frac{2^{k/4}}{\pi^{(k-2)/4} \, k^{1/2} } \bigg )^n \,
e^{-S_{r,k}} \ .
\label{crgen}
\eeqs
For cases where the number $k$ is obvious, we will sometimes suppress this
subscript in the notation.  


\section{Constraints from Limits on Baryon-Number-Violating Nucleon Decays} 
\label{pdecay_section}

In this section we analyze the constraints on fermion wavefunctions 
that can be derived from the experimental upper limits on
the rates for baryon-number-violating nucleon decays. We denote the relevant
BNV mass scale $M_{BNV}$ as $M_{Nd}$, where $Nd$ stands for ``nucleon decay''.
We assume that $M_{Nd}$ is large compared
with the highest gauge-symmetry breaking scale in the LRS model, namely $v_R$,
so that the effective Lagrangian is invariant under the LRS gauge
group, $G_{LRS}$.  

For the effective Lagrangian that is relevant for nucleon decays, we write 
\beq
{\cal L}^{(Nd)}_{eff}(x) = \sum_r c^{(Nd)}_r {\cal O}^{(Nd)}_r(x) + h.c. \ , 
\label{leff_pd}
\eeq
where $c^{(Nd)}_r$ are coefficients, and ${\cal O}^{(Nd)}_r(x)$ are the various
four-fermion operators.
Correspondingly, in the $d=(4+n)$-dimensional space, the effective Lagrangian
is
\beq
{\cal L}^{(Nd)}_{eff,4+n}(x,y) = 
\sum_r \kappa^{(Nd)}_r O^{(Nd)}_r(x,y) + h.c. \ . 
\label{leff_higherdim_pd}
\eeq

Four-fermion operators ${\cal O}^{(Nd)}_r$ in
${\cal L}^{(Nd)}_{eff}$ that contribute to nucleon decays in this LRS model and
are invariant under $G_{LRS}$ are listed below 
(where the unprimed and primed Roman indices are SU(2)$_L$ and
SU(2)$_R$ gauge indices, as defined above): 
\beqs
{\cal O}^{(Nd)}_{LL} &=& \epsilon_{\alpha\beta\gamma}
\epsilon_{ij} \epsilon_{km} 
[Q^{i \alpha \ T}_L C Q^{j \beta}_L]
[Q^{k \gamma \ T}_L C L^m_{\ell,L}] \cr\cr
&=& 2\epsilon_{\alpha\beta\gamma}[u^{\alpha \ T}_L C d^{\beta}_L]
\Big ( [u^{\gamma \ T}_L C \ell_L] - [d^{\gamma \ T}_L C \nu_{\ell,L}] \Big )
\cr\cr
&&
\label{op_ndec_LL}
\eeqs
\beqs
{\cal O}^{(Nd)}_{RR} &=& \epsilon_{\alpha\beta\gamma}
\epsilon_{i'j'} \epsilon_{k'm'} 
[Q^{i' \alpha \ T}_R C Q^{j' \beta}_R]
[Q^{k' \gamma \ T}_R C L^{m'}_{\ell,R}] \cr\cr
&=& 2\epsilon_{\alpha\beta\gamma}[u^{\alpha \ T}_R C d^{\beta}_R]
\Big ( [u^{\gamma \ T}_R C \ell_R] - [d^{\gamma \ T}_R C \nu_{\ell,R}] \Big )
\cr\cr
&&
\label{op_ndec_RR}
\eeqs
\beqs
{\cal O}^{(Nd)}_{LR} &=& \epsilon_{\alpha\beta\gamma}
\epsilon_{ij} \epsilon_{i'j'} 
[Q^{i \alpha \ T}_L C Q^{j \beta}_L]
[Q^{i' \gamma \ T}_R C L^{j'}_{\ell,R}] \cr\cr
&=& 2\epsilon_{\alpha\beta\gamma}[u^{\alpha \ T}_L C d^{\beta}_L]
\Big ( [u^{\gamma \ T}_R C \ell_R] - [d^{\gamma \ T}_R C \nu_{\ell,R}] \Big )
\cr\cr
&& 
\label{op_ndec_LR}
\eeqs
and
\beqs
{\cal O}^{(Nd)}_{RL} &=& \epsilon_{\alpha\beta\gamma}
\epsilon_{i'j'} \epsilon_{ij} 
[Q^{i' \alpha \ T}_R C Q^{j' \beta}_R]
[Q^{i \gamma \ T}_L C L^{j}_{\ell,L}] \cr\cr
&=& 2\epsilon_{\alpha\beta\gamma}[u^{\alpha \ T}_R C d^{\beta}_R]
\Big ( [u^{\gamma \ T}_L C \ell_L] - [d^{\gamma \ T}_L C \nu_{\ell,L}] \Big ) 
\ , \cr\cr
&&
\label{op_ndec_RL}
\eeqs
where $C$ is the Dirac charge conjugation matrix satisfying
$C \gamma_\mu C^{-1} = -(\gamma_\mu)^T$, $C=-C^T$; and $\epsilon_{\alpha\beta\gamma}$, 
$\epsilon_{ij}$, and $\epsilon_{i'j'}$ are totally antisymmetric SU(3)$_c$,
SU(2)$_L$, and SU(2)$_R$ tensors, respectively. 

To each of these operators ${\cal O}^{(Nd)}_r$ there corresponds an operator
$O^{(Nd)}_r$ in ${\cal L}^{(Nd)}_{eff,4+n}$. These are four-fermion
operators, and, as the $k=4$ special case of Eq. (\ref{kappagen}), we have
\beq
\kappa^{(Nd)}_r = \frac{\bar\kappa^{(Nd)}_r}{(M_{Nd})^{2+n}} \ .
\label{kappabarpd}
\eeq
The dependence of $\kappa^{(Nd)}_r$ on
the generational index of the lepton field that occurs in ${\cal O}^{(Nd)}_r$
is left implicit.  From the factorized form of fermion fields in
Eq. (\ref{psiform}), it follows that 
\beq
O^{(Nd)}_r(x,y) = U^{(Nd)}_r(x)V^{(Nd)}_r(y) \ , 
\label{opd_st}
\eeq
where $r=LL, \ RR, \ LR, \ RL$. 
To perform the integrals over $y$, we use the general integration formula given
as Eq. (A2) in \cite{bvd} and listed as Eq. (\ref{intform}) 
in the Appendix here. Carrying out the
integration over the $y$ components and using Eq. (\ref{a}) for the relevant
case $k=4$, we obtain the following results for the nonvanishing operators:
\beq
I^{(Nd)}_{LL} = b_4 \, \exp \bigg [ 
-\frac{3}{4}\|\eta_{Q_L}-\eta_{L_{\ell,L}}\|^2 \bigg ] 
\label{int_opLL_nucdec}
\eeq
\beq
I^{(Nd)}_{RR} = b_4 \, \exp \bigg [ 
-\frac{3}{4}\|\eta_{Q_R}-\eta_{L_{\ell,R}}\|^2 \bigg ] 
\label{int_opRR_nucdec}
\eeq
\beqs
I^{(Nd)}_{LR} &=& b_4 \, \exp \bigg [ -\frac{1}{4} \Big \{ 
  2\|\eta_{Q_L}-\eta_{Q_R}\|^2 +
  2\|\eta_{Q_L}-\eta_{L_{\ell_R}}\|^2 \cr\cr
  &+& 
 \|\eta_{Q_R}-\eta_{L_{\ell_R}}\|^2 \Big \} \bigg ] 
\label{int_opLR_nucdec}
\eeqs
and
\beqs
I^{(Nd)}_{RL} &=& b_4 \, \exp \bigg [ -\frac{1}{4} \Big \{ 
2\|\eta_{Q_R}-\eta_{Q_L}\|^2 + 
2\|\eta_{Q_R}-\eta_{L_{\ell_L}}\|^2 \cr\cr
&+& 
 \|\eta_{Q_L}-\eta_{L_{\ell_L}}\|^2 \Big \} \bigg ] 
\label{int_opRL_nucdec}
\eeqs
where $b_4 = (\pi^{-1/2} \mu )^n$, from the $k=4$ special case of
Eq. (\ref{bk}). It is convenient to write the integral $I^{(Nd)}_r$ in the 
form 
\beq
I^{(Nd)}_r \equiv b_4 \, e^{-S^{(Nd)}_r} \ , 
\label{bsi}
\eeq
where $S^{(Nd)}_r$ denotes the sum of squares of fermion wavefunction
separation distances (rescaled via multiplication by $\mu$ to be dimensionless)
in the argument of the exponent in $I^{(Nd)}_r$. Thus, for example, in the case
of $O^{(Nd)}_{LL}$, the sum in the exponent is
$S^{(Nd)}_{LL} = (3/4)\|\eta_{Q_L}-\eta_{L_{\ell,L}}\|^2$, and similarly for the other
$S^{(Nd)}_r$.  Then, as the special case of (\ref{crgen}) with 
$k=4$,
\beq
c^{(Nd)}_r = \frac{\bar\kappa^{(Nd)}_r}{(M_{Nd})^2} \, 
\bigg ( \frac{\mu}{\pi^{1/2} M_{Nd}} \bigg )^n \,  e^{-S^{(Nd)}_r} \ . 
\label{cr_nucleondecay}
\eeq

We use the experimental lower bound \cite{pdg} on the partial lifetime
$(\tau/B)_{N \to f.s.} = \Gamma_{N \to f.s.}^{-1}$ for a given nucleon
decay mode $N \to f.s.$ with branching ratio $B$ to a final state
denoted $f.s.$ to infer upper bounds on the magnitudes of the
$c^{(Nd)}_r$ coefficients.  The strongest lower bounds on these
partial lifetimes that are relevant here include $(\tau/B)_{p \to e^+
  \pi^0} > 1.6 \times 10^{34}$ yrs and $(\tau/B)_{p \to \mu^+ \pi^0} >
0.77 \times 10^{34}$ yrs \cite{abe17}.  The limits for the analogous
decays of neutrons are $(\tau/B)_{n \to e^+ \pi^-} > 0.53 \times
10^{34}$ yrs and $(\tau/B)_{n \to \mu^+ \pi^-} > 0.35 \times 10^{34}$
yrs \cite{abe17d}.  (These and other experimental limits quoted in
this paper are at the 90 \% confidence level.) Since we do not not
assume any cancellation between different terms
$c^{(Nd)}_r {\cal   O}^{(Nd)}_r$ occurring in ${\cal L}^{(Nd)}_{eff}$,
we impose the bounds from a given decay individually on each term that
contributes to it.  For given values of $\mu$, $M_{Nd}$, and the
dimensionless coefficients $\bar\kappa^{(Nd)}_r$, these constraints
are upper bounds on the integrals $I^{(Nd)}_r$ and hence lower bounds
on the the sums of squares of distances in $S^{(Nd)}_r$ for each
operator ${\cal O}^{(Nd)}_r$.  Our analysis of these lower bounds on
fermion separation distances in Ref. \cite{bvd} can be taken over,
with appropriate changes, for our present study; we refer the reader
to \cite{bvd} for the details. We find, for each $r$, $S^{(Nd)}_r >
(S^{(Nd)}_r)_{\rm min}$, where
\beqs
(S^{(Nd)}_r)_{\rm min} &=& 39 -\frac{n}{2} \ln \pi 
- 2 \ln \Big ( \frac{M_{Nd}}{10^4 \ {\rm TeV}} \Big ) \cr\cr
&-& n \, \ln \Big ( \frac{M_{Nd}}{\mu} \Big ) \ . 
\label{srminvalue}
\eeqs
The most direct bounds on fermion separation distances arises from the
contribution of the operators $O^{(Nd)}_{LL}$ and $O^{(Nd)}_{RR}$, since, for a
given $\ell$ ($=e$ or $\mu$), the integrals $I^{(Nd)}_{LL}$ and $I^{(Nd)}_{RR}$
each involve only one fermion separation distance, namely
$\|\eta_{Q_L}-\eta_{L_{\ell,L}}\|$ and $\|\eta_{Q_R}-\eta_{L_{\ell,R}}\|$,
respectively, for a given lepton generation $\ell=e$ or $\ell=\mu$.  In this
case, for the illustrative case of $n=2$ extra dimensions, 
we obtain the lower bound
\beqs
&& \| \eta_{Q_\chi} - \eta_{L_{\ell,\chi}}\|^2 > 50 
- \frac{8}{3}\ln \Big ( \frac{M_{Nd}}{10^4 \ {\rm TeV}} \Big ) 
- \frac{8}{3}\ln \Big ( \frac{M_{Nd}}{\mu} \Big ) \cr\cr
&& {\rm for} \ \chi = L, \ R \ {\rm and \ for} \ \ell = e, \ \mu \ . 
\label{opLLRR_pdec_constraint}
\eeqs

With the illustrative value $M_{Nd} = 10^4$ TeV, 
these are the inequalities $\| \eta_{Q_\chi} - \eta_{L_{\ell,\chi}}\| >
6.8$ for each of the four possibilities $\chi=L, \ R$ and $\ell=e, \ \mu$. A 
conservative solution to the coupled quadratic inequalities would require that
each of the relevant distances $\|\eta_{f_i}-\eta_{f_j}\|$ in
Eq. (\ref{opLLRR_pdec_constraint}) for both $\ell=e$ and $\ell=\mu$
would be larger than the square root of the right-hand side of
Eq. (\ref{srminvalue}):
\beqs
&& \Big \{ \ 
\|\eta_{Q_L}-\eta_{L_{\ell_L}} \|, \ 
\|\eta_{Q_R}-\eta_{L_{\ell_R}} \|, \ 
\|\eta_{Q_L}-\eta_{L_{\ell_R}} \|, \cr\cr
&& \|\eta_{Q_R}-\eta_{L_{\ell_L}} \| \ \Big \} 
> [(S^{(Nd)}_r)_{\rm min}]^{1/2} \ . 
\label{nucdec_quark_lepton_distances}
\eeqs
That is, this set of inequalities is sufficient, but not necessary, to satisfy
experimental constraints on the model from lower limits on partial lifetimes
for nucleon decays. 


\section{$n - \bar n$ Oscillations and Dinucleon Decays} 
\label{nnbar_section}

In this section we analyze $n - \bar n$ oscillations and the resultant
$\Delta B=-2$ dinucleon decays in this extra-dimensional LRS model. We
refer the reader to Refs. \cite{nnb02} and \cite{bvd} for relevant
background; here we will review this background briefly. We consider a
general theory in which baryon-number violating physics can produce
$n-\bar n$ transitions.  We denote the relevant low-energy effective
Lagrangian in 4D as ${\cal L}^{(n\bar n)}_{eff}$, and the transition
matrix element as $|\delta m| = |\langle \bar n |{\cal L}^{(n\bar
  n)}_{eff} | n \rangle|$.  In (field-free) vacuum, an initial state
which is $|n\rangle$ at time $t=0$ has a nonzero probability to be an
$|\bar n\rangle$ state at a later time $t > 0$.  This probability is
given by
$P(n(t)=\bar n) = |\langle \bar n|n(t) \rangle|^2 = [\sin^2(t/\tau_{n \bar n})]e^{-t/\tau_n}$,
where $\tau_{n \bar n} =
1/|\delta m|$ and $\tau_n$ is the mean life of the neutron. The
current direct limit on $\tau_{n \bar n}$, from a reactor experiment
at the Institut Laue-Langevin (ILL) in Grenoble, is $\tau_{n \bar n}
\ge 0.86 \times 10^8$ sec, i.e., $|\delta m| < 0.77 \times 10^{-29}$
MeV \cite{ill}. Because of the nonvanishing $n-\bar n$ transition
amplitude, the physical eigenstate for the neutron state in matter has
a small component of $\bar n$, i.e.,
$|n\rangle_{\rm phys.} = \cos\theta_{n\bar n} |n\rangle + \sin\theta_{n \bar n} |\bar n\rangle$,
with $|\theta_{n \bar n}| \ll 1$. In turn, this leads to
annihilation with an adjacent neutron or proton, and hence to $\Delta
B=-2$ decays to nonbaryonic final states, predominantly involving
pions. Experiments have searched for the resultant matter instability
due to these dinucleon decays and have set lower limits on the matter
instability (m.i.) lifetime, $\tau_{\rm m.i.}$. This lifetime is
related to $\tau_{n \bar n}$ by the formula $\tau_{m.i.} = R \,
\tau_{n \bar n}^2$, where $R \sim O(10^2)$ MeV, or equivalently, $R
\simeq 10^{23}$ sec$^{-1}$, depending on the nucleus. The best current
limit on matter instability is from the SuperKamiokande (SK) water
Cherenkov experiment \cite{sk_nnbar}, namely $\tau_{m.i.} > 1.9 \times
10^{32}$ yr. Using the value $R \simeq 0.52 \times 10^{23}$ sec$^{-1}$
for the ${}^{16}$O nuclei in the water, Ref. \cite{sk_nnbar} obtained
the lower bound $\tau_{n \bar n} > 2.7 \times 10^8$, or equivalently,
\beq
|\delta m | < 2.4 \times 10^{-30} \ {\rm MeV} . 
\label{deltam_limit}
\eeq

As mentioned above, we shall analyze $n-\bar n$ oscillations in this theory by
writing down the relevant $G_{LRS}$-invariant operators, which are six-quark
operators multiplied by $(\Delta_R)^\dagger$, and then focusing on the
resultant $|\Delta B|=2$ six-quark operators resulting from the VEV of
$(\Delta_R)^\dagger$. The effective Lagrangian (in four-dimensional spacetime)
that mediates $n-\bar n$ oscillations is a sum of six-quark operators,
\beq
{\cal L}^{(n \bar n)}_{eff}(x) = \sum_r c^{(n \bar n)}_r \, 
{\cal O}^{(n \bar n)}_r(x) + h.c. \ . 
\label{leff_nnbar}
\eeq
The corresponding Lagrangian in the $(4+n)$-dimensional space is
\beq
{\cal L}^{(n \bar n)}_{eff,4+n}(x,y) = 
\sum_r \kappa^{(n \bar n)}_r O^{(n \bar n)}_r(x,y) + h.c. \ . 
\label{leff_higherdim_nnbar}
\eeq
We find, for the set ${\cal O}^{(n \bar n)}_r$, the operators 
\begin{widetext}
\beq
{\cal O}^{(n \bar n)}_1 = (T_s)_{\alpha\beta\gamma\delta\rho\sigma} \, 
(\epsilon_{i'k'}\epsilon_{j'm'} + \epsilon_{j'k'}\epsilon_{i'm'} ) 
(\epsilon_{p'r'}\epsilon_{q's'} + \epsilon_{q'r'}\epsilon_{p's'} )  
\,
[Q_R^{i' \alpha T} C Q_R^{j' \beta}]
[Q_R^{k' \gamma T} C Q_R^{m' \delta}]
[Q_R^{p' \rho T} C Q_R^{q' \sigma} ] \, (\Delta_R^\dagger)^{r's'} 
\label{op_nnbar_1}
\eeq
\beq
{\cal O}^{(n \bar n)}_2 = (T_a)_{\alpha\beta\gamma\delta\rho\sigma} \, 
\epsilon_{i' j'} \epsilon_{k' m'} \, 
(\epsilon_{p'r'}\epsilon_{q's'} + \epsilon_{q'r'}\epsilon_{p's'} ) 
\, 
[Q_R^{i' \alpha T} C Q_R^{j' \beta}]
[Q_R^{k' \gamma T} C Q_R^{m' \delta}]
[Q_R^{p' \rho T} C Q_R^{q' \sigma} ] \, (\Delta_R^\dagger)^{r's'} 
\label{op_nnbar_2}
\eeq
\beq
{\cal O}^{(n \bar n)}_3 = (T_a)_{\alpha\beta\gamma\delta\rho\sigma} \, 
\epsilon_{i j} \epsilon_{k' m'} \, 
(\epsilon_{p'r'}\epsilon_{q's'} + \epsilon_{q'r'}\epsilon_{p's'} ) 
\, 
[Q_L^{i \alpha T} C Q_L^{j \beta}]
[Q_R^{k' \gamma T} C Q_R^{m' \delta}]
[Q_R^{p' \rho T} C Q_R^{q' \sigma} ] \, (\Delta_R^\dagger)^{r's'} 
\label{op_nnbar_3}
\eeq
\beq
{\cal O}^{(n \bar n)}_4 = (T_a)_{\alpha\beta\gamma\delta\rho\sigma} \, 
\epsilon_{i j} \epsilon_{k m} \, 
(\epsilon_{p'r'}\epsilon_{q's'} + \epsilon_{q'r'}\epsilon_{p's'} ) 
\, 
[Q_L^{i \alpha T} C Q_L^{j \beta}]
[Q_L^{k \gamma T} C Q_L^{m \delta}]
[Q_R^{p' \rho T} C Q_R^{q' \sigma} ] \, (\Delta_R^\dagger)^{r's'} 
\label{op_nnbar_4}
\eeq
\beq
{\cal O}^{(n \bar n)}_5 = (T_s)_{\alpha\beta\gamma\delta\rho\sigma}
(\epsilon_{ik}\epsilon_{jm} + \epsilon_{jk}\epsilon_{im} ) 
(\epsilon_{p'r'}\epsilon_{q's'} + \epsilon_{q'r'}\epsilon_{p's'} ) 
\, 
[Q_L^{i \alpha T} C Q_L^{j \beta}]
[Q_L^{k \gamma T} C Q_L^{m \delta}]
[Q_R^{p' \rho T} C Q_R^{q' \sigma} ] \, (\Delta_R^\dagger)^{r's'} 
\label{op_nnbar_5}
\eeq
\end{widetext}
where the SU(3)$_c$ color tensors are 
\beqs
(T_s)_{\alpha \beta \gamma \delta \rho \sigma} &=&
\epsilon_{\rho \alpha \gamma}\epsilon_{\sigma \beta \delta} +
\epsilon_{\sigma \alpha \gamma}\epsilon_{\rho \beta \delta} \cr\cr
&+& 
\epsilon_{\rho \beta \gamma}\epsilon_{\sigma \alpha \delta} +
\epsilon_{\sigma \beta \gamma}\epsilon_{\rho \alpha \delta}
\label{ts}
\eeqs
and
\beq
(T_a)_{\alpha \beta \gamma \delta \rho \sigma} =
\epsilon_{\rho \alpha \beta}\epsilon_{\sigma \gamma \delta} +
\epsilon_{\sigma \alpha \beta}\epsilon_{\rho \gamma \delta} \ . 
\label{ta}
\eeq

To obtain the six-quark operators that mediate $n-\bar n$ transitions, we
replace the $\Delta_R$ field by its VEV, $v_R$. To each of these 
$n-\bar n$ transition operators ${\cal O}^{(n\bar n)}_r$ there corresponds an 
operator $O^{(n \bar n)}_r$ in ${\cal L}^{(Nd)}_{eff,4+n}$.  We have
\beq
\kappa^{(n \bar n)}_r = \frac{\bar\kappa^{(n \bar n)}_r}
{(M_{n \bar n})^{6 + (5n/2)} } \ .
\label{kappabarpd2}
\eeq
To each of these operators there is a corresponding $V^{(n \bar n)}_r$ 
function; for example, 
\beqs
&& V^{(n \bar n)}_1 = V^{(n \bar n)}_2 = A^6 \exp 
\Big [ - 6\|\eta-\eta_{Q_R}\|^2 \Big ] \ , \cr\cr
&&
\label{chiprod_op12_nnbar}
\eeqs
and so forth for the others. The integrals of these
functions over the extra $n$ dimensions comprise two classes. The integration
of the $V^{(n \bar n)}_r$ functions for the operators ${\cal O}^{(n \bar n)}_r$
with $r=1,2$ are the same, defining class $C^{(n \bar n)}_{1s}$, where the
subscript $s$ is appended to distinguish this and the other classes from the
classes calculated in terms of the $G_{SM}$-based low-energy effective field
theory in \cite{nnb02,bvd}:
\beq
I^{(n \bar n)}_{C_{1s}} = b_6 
\label{int_class1s_nnbar}
\eeq
where $b_6 = (2 \cdot 3^{-1/2} \, \pi^{-1} \mu ^2)^n$
from the $k=6$ special case of Eq. (\ref{bk}) and
$I^{(n \bar n)}_{C_k} \equiv I_{C^{(n \bar n)}_k}$. 
The integrals of the operators ${\cal O}^{(n \bar n)}_r$ with $r=3,4,5$ 
are equal and yield a second class, 
\beq
I^{(n \bar n)}_{C_{2s}} = b_6 \, \exp \Big [ 
-\frac{4}{3}\|\eta_{Q_L}-\eta_{Q_R} \|^2 \Big ] \ . 
\label{int_class2s_nnbar}
\eeq

From the special case of Eq. (\ref{crgen}) with $k=6$, together with 
Eq. (\ref{m_nnb_vr}), it follows that 
\beq
c^{(n \bar n)}_r = \frac{\bar\kappa^{(n \bar n)}_r}{v_R^5} \,
\bigg ( \frac{2\mu^2}{3^{1/2}\pi v_R^2} \bigg )^n \,
e^{-S^{(n \bar n)}_r} \ , 
\label{cr_nnbar}
\eeq
where 
\beq
S^{(n \bar n)}_r = 0 \quad {\rm for} \ r=1,2 
\label{sr12}
\eeq
and
\beq
S^{(n \bar n)}_r = \frac{4}{3} \|\eta_{Q_L}-\eta_{Q_R} \|^2 
\quad {\rm for} \ r=3,4,5 \ .  
\label{sr345}
\eeq
An important result from this calculation is that because $S^{(n \bar n)}_r=0$
for $r=1,2$, there is no exponential wavefunction suppression from the
integration over the $n$ extra dimensions for $O^{(n \bar n)}_r$ with
$r=1,2$.

Then
\beqs
|\delta m| &=& \frac{1}{v_R^5} \, 
\Big ( \frac{\mu}{v_R} \Big )^{2n}
\, \Big (\frac{2}{3^{1/2} \pi} \Big )^n \cr\cr
&\times& 
\Big |\sum_r \bar\kappa^{(n \bar n)}_r \, 
e^{-S^{(n \bar n)}_r} \, \langle \bar n | {\cal O}^{(n \bar n)}_r | n\rangle 
\Big | \ . 
\label{deltam_calc}
\eeqs
The dominant contribution to $|\delta m|$ comes from the operators
${\cal O}^{(n \bar n)}_r$ with $r=1,2$ (provided that the coefficients
$\bar\kappa^{(n \bar n)}_r$ with $r=1,2$ are not negligibly small),
since $S^{(n \bar n)}_r=0$ for $r=1,2$, so these operators do not
incur any exponential suppression factors from the integration over
the extra dimensions.  The matrix elements $\langle \bar n | {\cal
  O}^{(n \bar n)}_r | n \rangle$ have dimensions of $({\rm mass})^6$,
and since they are determined by hadronic physics, one expects on
general grounds that they are $\sim \Lambda_{QCD}^6$, where, as above,
$\Lambda_{QCD} \simeq 0.25$ GeV. This expectation is confirmed by
quantitative calculations \cite{nnb82,nnb84,nnblgt}.  Taking
$\bar\kappa^{(n \bar n)}_r \sim O(1)$ for $r=1,2$ and the illustrative
value $n=2$ extra dimensions, and requiring that $|\delta m|$ must be
less than the experimental upper bound (\ref{deltam_limit}), we then
derive the following lower bound on $M_{n \bar n}=v_R$:
\beqs
 v_R &>& (1 \times 10^3 \ {\rm TeV}) \Big ( 
 \frac{\tau_{n \bar n}}{2.7 \times 10^8 \ {\rm sec}} \Big )^{1/9} \cr\cr
 &\times&
\Big ( \frac{\mu}{3 \times 10^3 \ {\rm TeV} } \Big )^{4/9} 
\bigg ( \frac{|\langle \bar n | {\cal O}^{(n \bar n)}_4 | n\rangle | }
      {\Lambda_{QCD}^6} \bigg )^{1/9}  \ . \cr\cr
      &&
\label{vr_min}
\eeqs

Thus, our analysis shows that, while it is easy to suppress $\Delta B=-1$
nucleon decay far below observable levels in this model by making the fermion
wavefunction separation distances in Eq. (\ref{nucdec_quark_lepton_distances})
sufficiently large, this does not suppress the $|\Delta B|=2$ \ $n-\bar n$
oscillations, which can occur at a level comparable with current experimental limits.
We have used this fact to deduce the lower bound (\ref{vr_min}) on
$v_R$ and hence the scale of $|\Delta B|=2$ baryon number violation in this model.
A similar comment applies to $\Delta B=-2$ dinucleon decays (occurring primarily
to multipion final states), since these are induced by the fundamental
$n - \bar n$ oscillations. 

It is of interest to compare our new results for the extra-dimensional
LRS model with the results that were previously obtained in
Ref. \cite{nnb02} and studied further in \cite{bvd} for an
extra-dimensional model that used a Standard-Model low-energy field
theory.  A striking feature that is common to both of these types of
models is that although one can easily arrange the fermion
wavefunction separation distances to suppress nucleon decays, this
does not suppress $n - \bar n$ oscillations.  A basic difference
between the model used in Refs. \cite{nnb02,bvd} and the present LRS
model is that in the SM effective field theory framework of
\cite{nnb02,bvd}, baryon number is a global symmetry, while in the LRS
model, $B$ and $L$ are gauged via the U(1)$_{B-L}$ symmetry, and the
VEV of the $\Delta_R$ field spontaneously breaks $B$ by 2 units in
processes for which $\Delta L=0$.  Hence, while the SM Higgs VEV
preserves $B$ (and $L$), here the scale of baryon number violation is
set by $v_R$, as given in Eq. (\ref{m_nnb_vr}).  We recall the
corresponding limit from Ref. \cite{nnb02} (updated in \cite{bvd} with
the newer limit on $\tau_{m.i.}$ from the SK experiment
\cite{sk_nnbar}), namely
\beqs
M_{n \bar n} &>& (44 \ {\rm TeV}) \Big (
\frac{\tau_{n \bar n}}{2.7 \times 10^8 \ {\rm sec}} \Big )^{1/9} \cr\cr
&\times&
\Big ( \frac{\mu}{3 \times 10^3 \ {\rm TeV} } \Big )^{4/9} 
\bigg ( \frac{|\langle \bar n | {\cal O}^{(n \bar n)}_4 | n\rangle | }
      {\Lambda_{QCD}^6} \bigg )^{1/9}   \cr\cr
      && {\rm for \ SMEFT} \ .
\label{M_nnbar_min_smeft}
\eeqs
The main reason why the lower bound on $M_{n \bar n}=v_R$ in
Eq. (\ref{vr_min}) is substantially higher than the lower bound on
$M_{n \bar n}$ in Eq. (\ref{M_nnbar_min_smeft}) is that all of the
integrals of six-quark operators in the extra dimensions in the model
of Refs. \cite{nnb02,bvd} involved exponential suppression factors,
whereas, in contrast, here, $S^{(n \bar n)}_r=0$ for $r=1,2$, so the
integrals of these operators $O^{(n \bar n)}_r$ over the extra
dimensions do not produce any exponential suppression factors.


\section{Conclusions}
\label{conclusion_section} 

In this paper we have studied $n-\bar n$ oscillations in a
left-right-symmetric model in which Standard Model fermions have
localized wavefunctions in extra dimensions.  We have shown that in
this extra-dimensional LRS model, even with fermion wavefunction
positions chosen so as to render the rates for baryon-violating
nucleon decays much smaller than experimental limits, $n-\bar n$
oscillations can occur at rates comparable to current bounds. Thus,
this feature is common to both the present LRS model and the model
with a SM low-energy effective field theory studied in
\cite{nnb02,bvd}.  An interesting difference between these models that
we find is that certain six-quark operators in the LRS model are not
suppressed by exponential factors resulting from the integration over
the extra dimensions, in contrast to the SMEFT model of
Refs. \cite{nnb02,bvd}, where this integration yields exponential
suppression factors for all six-quark operators.  These findings
provide further motivation for new experimental searches for $n-\bar
n$ oscillations. In the future, one may look forward to such
experiments using a neutron beam at the European Spallation Source
\cite{nnbar_physrep} and searching for resultant matter instability in
the water Cherenkov detector in Hyper-Kamiokande \cite{hk} and the
liquid argon detector in the Deep Underground Neutrino Experiment,
DUNE \cite{dune,barrow}.


\acknowledgments

This research was supported in part by the NSF Grant NSF-PHY-1915093. 


\appendix
\section{An Integral Identity}
\label{integral_appendix}

Let $\eta$ be an $n$-dimensional vector $\eta \in
{\mathbb R}^n$ with components $\eta_j$, $j=1,...,n$ and
norm $\| \eta \| = [\sum_{j=1}^n \eta_j^2]^{1/2}$ and let 
$[\prod_{j=1}^n \int_{-\infty}^\infty d\eta_j]\, F(\eta)
\equiv \int d^n \eta \, F(\eta)$.  An integral identity 
\cite{bvd} that we use is 
\begin{widetext}
\beq
\int d^n \eta \, \exp
\Big [-\sum_{i=1}^m a_i\|\eta-\eta_{f_i}\|^2 \Big ]
= \bigg [ \frac{\pi}{\sum_{i=1}^m a_i} \bigg ]^{n/2} \,
\exp\Bigg [ \frac{-\sum_{j,k=1; \ j < k}^m \, a_j a_k
\|\eta_{f_j}-\eta_{f_k}\|^2}{\sum_{s=1}^m a_s} \Bigg ] \ . 
\label{intform}
\eeq
\end{widetext}
Note that if $m=1$, then the argument of the exponent
is zero and the right-hand side of (\ref{intform}) is simply $(\pi/a_1)^{n/2}$.




\begin{thebibliography}{99}
\bibitem{su2instantons}
  As analyzed in \cite{hooft}, SU(2)$_L$ instantons in the SM break $B$ and $L$
  (conserving $B-L$), but this has a negligibly small effect at temperatures
  much smaller than the electroweak symmetry breaking scale.
\end{thebibliography}
\end{document}